\documentclass[
 reprint,
 amsmath,amssymb,superscriptaddress,
 aps,
 prl,
groupedaddress
]{revtex4-1}

\usepackage{graphicx}
\usepackage{dcolumn}
\usepackage{bm}
\usepackage{color}
\usepackage{ulem}
\usepackage{float}
\usepackage{textcomp}
\restylefloat{table}
\usepackage{verbatim}

\begin{document}

\title{Anisotropic composite fermions and fractional quantum Hall effect}
\date{\today}

\author{M. A.\ Mueed}
\author{D.\ Kamburov}
\author{S.\ Hasdemir}
\author{L. N.\ Pfeiffer}
\author{K. W.\ West}
\author{K. W.\ Baldwin}
\author{M.\ Shayegan}
\affiliation{Department of Electrical Engineering, Princeton University, Princeton, New Jersey 08544, USA}

\begin{abstract}
We study the role of anisotropy on the transport properties of composite fermions near Landau level filling factor $\nu=1/2$ in two-dimensional holes confined to a GaAs quantum well. By applying a parallel magnetic field, we tune the composite fermion Fermi sea anisotropy and monitor the relative change of the transport scattering time at $\nu=1/2$ along the principal directions. Interpreted in a simple Drude model, our results suggest that the scattering time is longer along the longitudinal direction of the composite fermion Fermi sea. Furthermore, the measured energy gap for the fractional quantum Hall state at $\nu=2/3$ decreases when anisotropy becomes significant. The decrease, however, might partly stem from the charge distribution becoming bilayer-like at very large parallel magnetic fields. 
\end{abstract} 

\maketitle
The rich many-body physics of two-dimensional (2D) carriers inherent in the fractional quantum Hall effect (FQHE), even three decades after its discovery, continues to spur exciting research \cite{Tsui.PRL.1982,Laughlin.PRL.1983,Jain.2007}. There has been a recent surge of interest in studies, both experimental and theoretical, of anisotropy in interacting electron systems and in particular the FQHE \cite{Mueed1.PRL.2015,Xia.NatPhy.2011,Gokmen.NatPhy.2010,Liu.PRB.2013, Mulligan.PRB.2010,Bo.PRB.2012,Qiu.PRB.2012,Wang.PRB.2012,Wang.PRB.2012,Papic.PRB.2013,Yang.PRB.2013,Balram.preprint, Kamburov.PRL.2013,Kamburov.PRB.2014,Fradkin.ARCM.2010,Haldane.PRL.2011,Balagurov.PRB.2000}.  FQHE states, associated with Laughlin's wave function \cite{Laughlin.PRL.1983}, have been historically considered to be isotropic and rotationally invariant. However, in light of a revelation by Haldane \cite{Haldane.PRL.2011}, they are understood to also possess a geometric degree of freedom intimately linked to the underlying anisotropy of the 2D system. Of fundamental interest is how such anisotropy affects properties of the FQHE states and the composite fermions (CFs), quasi-particles which provide an elegant description of the FQHE \cite{Jain.2007,Jain.PRL.1989,Halperin.PRB.1993}.

Here we address this question through transport measurements on a 2D system which is rendered anisotropic via the application of a large in-plane magnetic field ($B_{||}$). In a strictly 2D system with zero thickness, the in-plane motion of the carriers is unaffected by $B_{||}$. However, for quasi-2D systems with finite width, such as electrons in a quantum well (QW), $B_{||}$ can couple to electrons' out-of-plane motion, thus also affecting their in-plane motion. Because of such coupling, the electron Fermi contour becomes anisotropic. When subjected to $B_{||}$, CFs, too, show qualitatively similar behaviour. In Fig. 1, we focus on three aspects of this $B_{||}$-induced anisotropy: (i) the anisotropy of CFs' Fermi contour near Landau level filling $\nu=1/2$, (ii) the anisotropy of CFs' resistivity at $\nu=1/2$, and (iii) the observation of a significant reduction of the energy gap of the nearby $\nu=2/3$ FQHE which we discuss in light of anisotropy as well as a possible $B_{||}$-induced single-layer to bilayer transition of the charge distribution. This combination of data sheds light on the CF and FQHE anisotropy, and provides valuable input for future work.

\begin{figure}
\includegraphics[width=.45\textwidth]{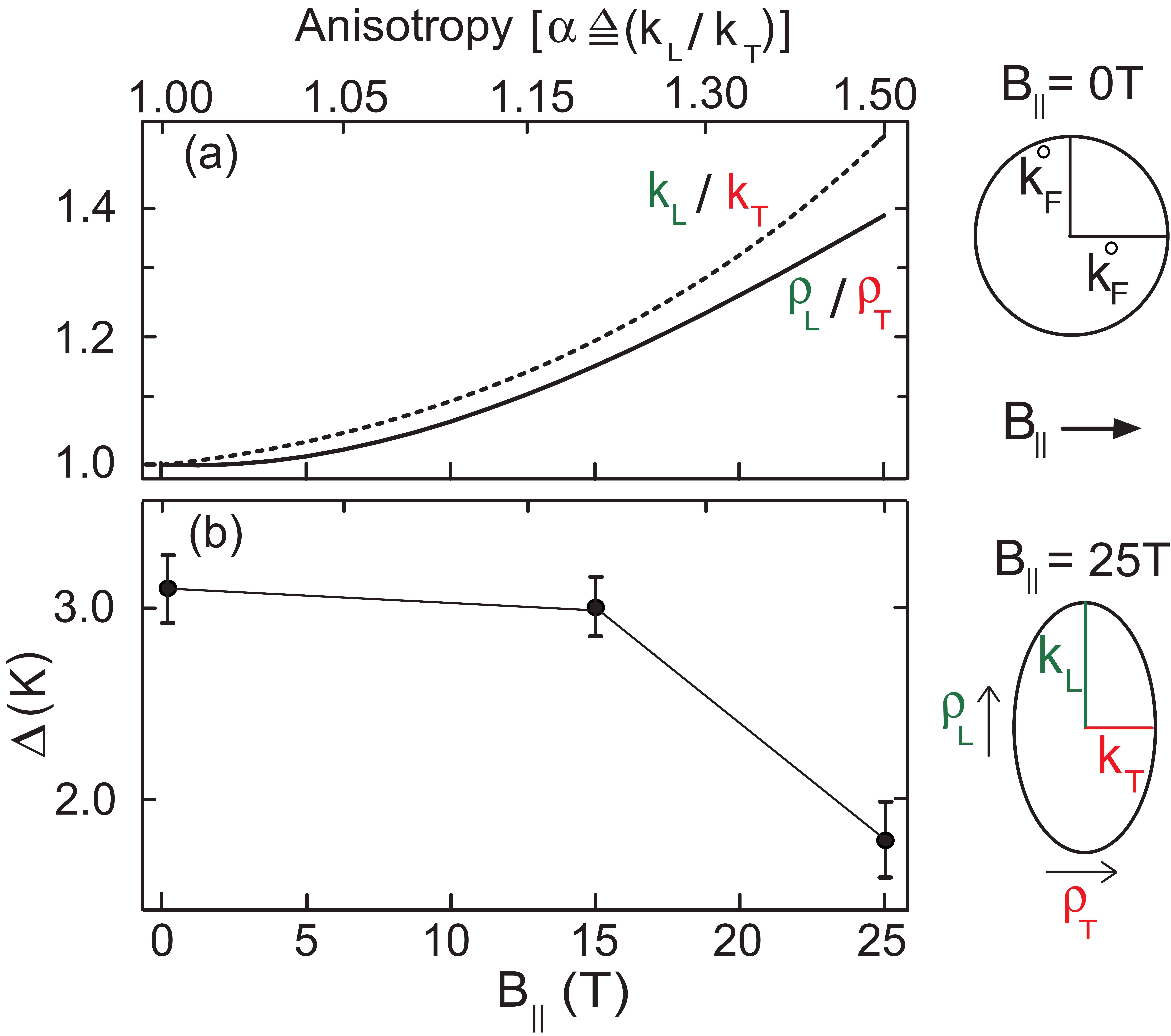}
\caption{\label{fig:Fig1} (color online) \textit{Summary of our results}. On the right, we show the evolution of CF Fermi sea in the presence of $B_{||}$. Because of the finite layer thickness of the system, $B_{||}$ couples to the out of plane motion of CFs and distorts their $circular$ Fermi contour into an $elliptical$ shape. The Fermi contour diameter shrinks along $B_{||}$ and gets elongated in the perpendicular direction. Throughout the paper, we denote the measured quantities along the transverse and longitudinal directions of the CF Fermi contour with subscripts $T$ and $L$, respectively. Note that these directions correspond to parallel and perpendicular to the direction of $B_{||}$. (a) The measured ratios of CFs' resistivities ($\rho_L/\rho_T$) and Fermi wave vectors ($k_L/k_T$) as a function of $B_{||}$. (b) The $\nu=2/3$ FQHE energy gap ($\Delta$) vs $B_{||}$. In the upper horizontal axis, we also mark the corresponding values of the anisotropy factor $\alpha$, which we define as the measured ($k_L/k_T$) for CFs (see text).}
\end{figure}

We studied CFs of a 2D hole system (2DHS) grown via molecular beam epitaxy. The 2DHS is confined to a 17.5-nm-wide, symmetric GaAs (001) QW which is located 136 nm below the surface and is flanked on each side by 95-nm-thick Al$_{0.24}$Ga$_{0.76}$As layers and C $\delta$-doped layers. It has density $p$ $\simeq 1.43\times10^{11}$ cm$^{-2}$, and low temperature mobility $\simeq 10^{6}$ cm$^{2}$/Vs. We fabricated two L-shaped Hall bar samples with the perpendicular arms oriented along [110] and $[\overline{1}10]$. One sample (Fig. 2(a)) has a periodic grating of negative electron-beam resist patterned on its surface to induce a potential modulation for the 2D carriers. The other sample, as shown in Fig. 2(b), is unpatterned. 
We recorded, at $T=0.3$ K, the resistivity along the two arms in purely perpendicular and also in tilted magnetic fields, with $\theta$ denoting the tilt angle. The samples were rotated around $[\overline{1}10]$ so that $B_{||}$ was always parallel to [110] (Fig. 2). This configuration orients the longitudinal and transverse axes of the elliptical Fermi contour along $[\overline{1}10]$ and [110], respectively.

\begin{figure}
\includegraphics[width=.47\textwidth]{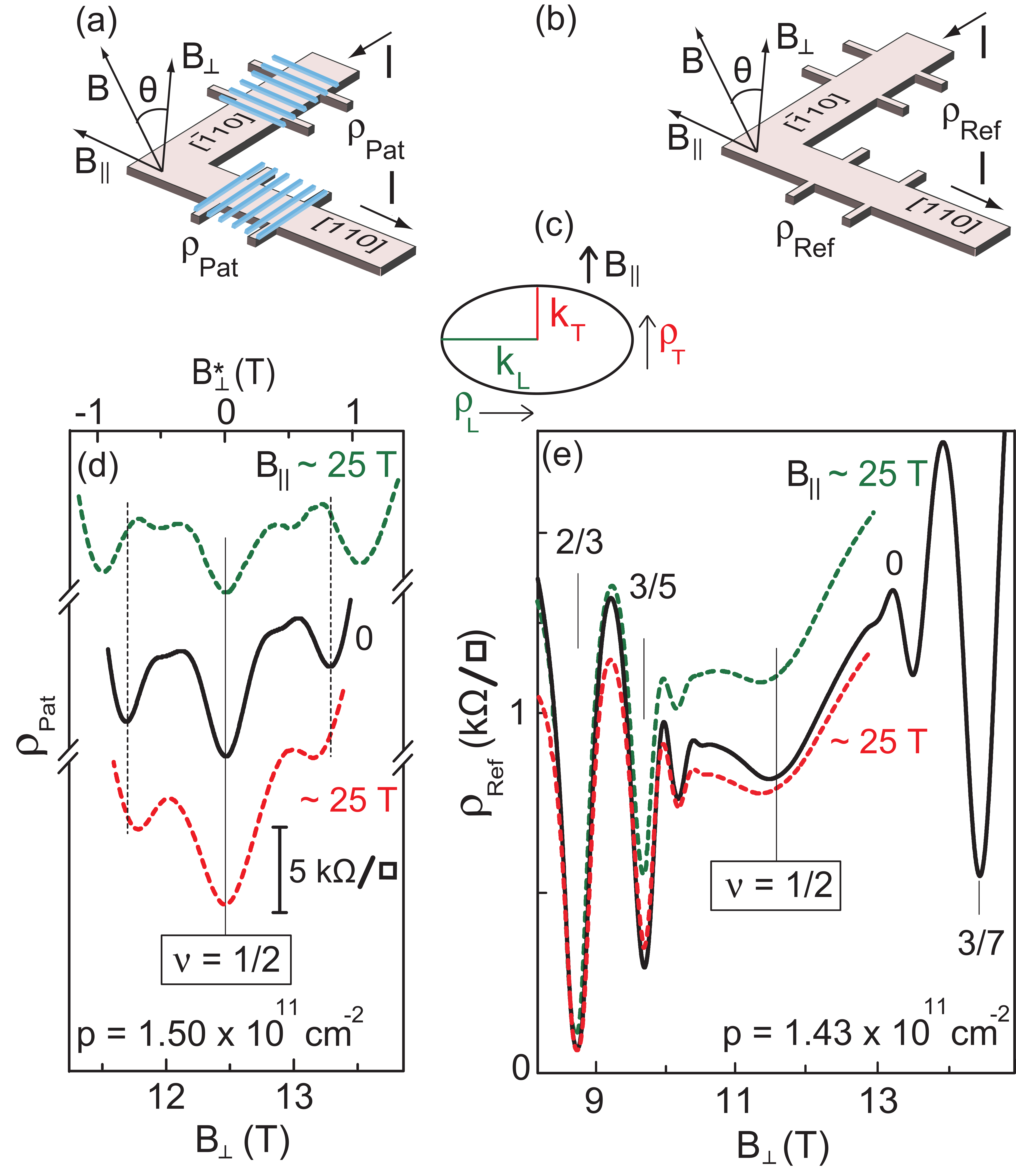}
\caption{\label{fig:Fig1} (color online) (a) Patterned and (b) unpatterned (Reference) L-shaped Hall bar samples. (c) Elliptical Fermi contour of CFs in the presence of $B_{||}$. (d) Magnetoresistivity traces of our 2DHS sample subjected to a weak, strain-induced, unidirectional periodic potential modulation \cite{Kamburov.PRL.2013}. The resistivity minima on the flanks of $\nu=1/2$ signal the geometric resonance of CFs' cyclotron orbit with the modulation period and provide a measure of their Fermi wave vector $k_F$. The minima from the black trace, taken at $B_{||}=0$, match the expected positions (marked by dotted vertical lines) of the primary commensurability minima of CFs if their Fermi contour were $circular$ \cite{footnote2,Kamburov.PRL.2014}. The green and red traces, for which $B_{||}\sim25$ T at $\nu=1/2$, probe $k_F$ along the longitudinal and transverse axes of the $elliptical$ Fermi contour, respectively. Traces are shifted vertically for clarity. (e) The corresponding magnetoresistivity traces from the unpatterned sample. Unlike in (d), the traces are not shifted.}
\end{figure}

Before presenting the experimental data in detail, we briefly discuss CFs, exotic quasi-particles each composed of one charged particle (electron or hole) and an even number of flux quanta. The CF concept has been very successful in explaining the many-body physics in 2D at large perpendicular magnetic fields ($B_{\perp}$) \cite{Jain.2007, Jain.PRL.1989, Halperin.PRB.1993}. Thanks to the flux attachment which cancels the external magnetic field at a half-filled Landau level, one of CFs' remarkable properties is that, at $\nu=1/2$, they behave as if they are at $B_{\perp}=0$. Away from $\nu=1/2$, CFs feel the effective magnetic field $B^*_{\perp}=B_{\perp}-B_{\perp,1/2}$, where $B_{\perp,1/2}$ is the field at $\nu=1/2$. In the limit of small $B^*_{\perp}$, CFs occupy a well-defined Fermi sea \cite{Halperin.PRB.1993, Willett.PRL.1993, Kang.PRL.1993, Goldman.PRL.1994, Smet.PRL.1998, Kamburov.PRL.2013, Kamburov.PRB.2014} with a circular Fermi contour if the system is isotropic. Moreover, qualitatively similar to their zero-field counterpart particles, the application of $B_{||}$ induces anisotropy in the Fermi sea of CFs (see Fig. 1) by coupling to their out-of-plane motion through the finite  thickness of the charge distribution \cite{Kamburov.PRL.2013,Kamburov.PRB.2014}.

Figure 2(d) shows the geometric resonance features of $\nu=1/2$ CFs as the 2DHS is subjected to a lateral density modulation stemming from the periodic surface grating. On both sides of $\nu=1/2$, we observe resistance minima signaling the commensurability of CFs' cyclotron orbit diameter with the modulation period. Positions of these minima measured relative to $B_{\perp,1/2}$ are directly proportional the CFs' Fermi wave vector ($k_F$). The commensurability features of the black trace, taken at $B_{||}=0$, are consistent with the dotted vertical lines based on the wave vector of a circular Fermi contour ($k_F^o$) (see Fig. 1) and full spin polarization \cite{Roland book,Kamburov.PRL.2014,footnote2,footnote2_1,Kukushkin.PRL.1999,Liu.PRB.2014}. The green and red traces probe the elliptical Fermi contour of CFs at $B_{||}\simeq25$ T along its longitudinal and transverse directions, respectively. Compared to the black trace, the minima in the green trace move away from $\nu=1/2$, while in the red trace they move closer \cite{footnote3}. This indicates that the Fermi contour becomes elongated in the longitudinal direction but shrinks in the transverse direction under $B_{||}$, as illustrated in Fig. 1. 

Since the external density modulation in a patterned sample introduces additional scattering for the CFs, we use traces from the unpatterned sample to determine accurate values of resistivity ($\rho$) for the $\nu=1/2$ CFs. Figure 2(e) shows such traces. Compared to the $B_{||}=0$ case, the resistivity at $\nu=1/2$ in the longitudinal direction ($\rho_L$) increases, while it decreases in the transverse direction ($\rho_T$). In Fig. 3, we plot the independently measured quantities $\rho_L$,  $\rho_T$, $k_L$ and $k_T$ for CFs as a function of $B_{||}$; these are all normalized to their respective $B_{||}=0$ values \cite{footnote1}. We find that while both $\rho_L$ and $k_L$ increase with increasing $B_{||}$, $\rho_T$ and $k_T$ decrease.


\begin{figure}
\includegraphics[width=.40\textwidth]{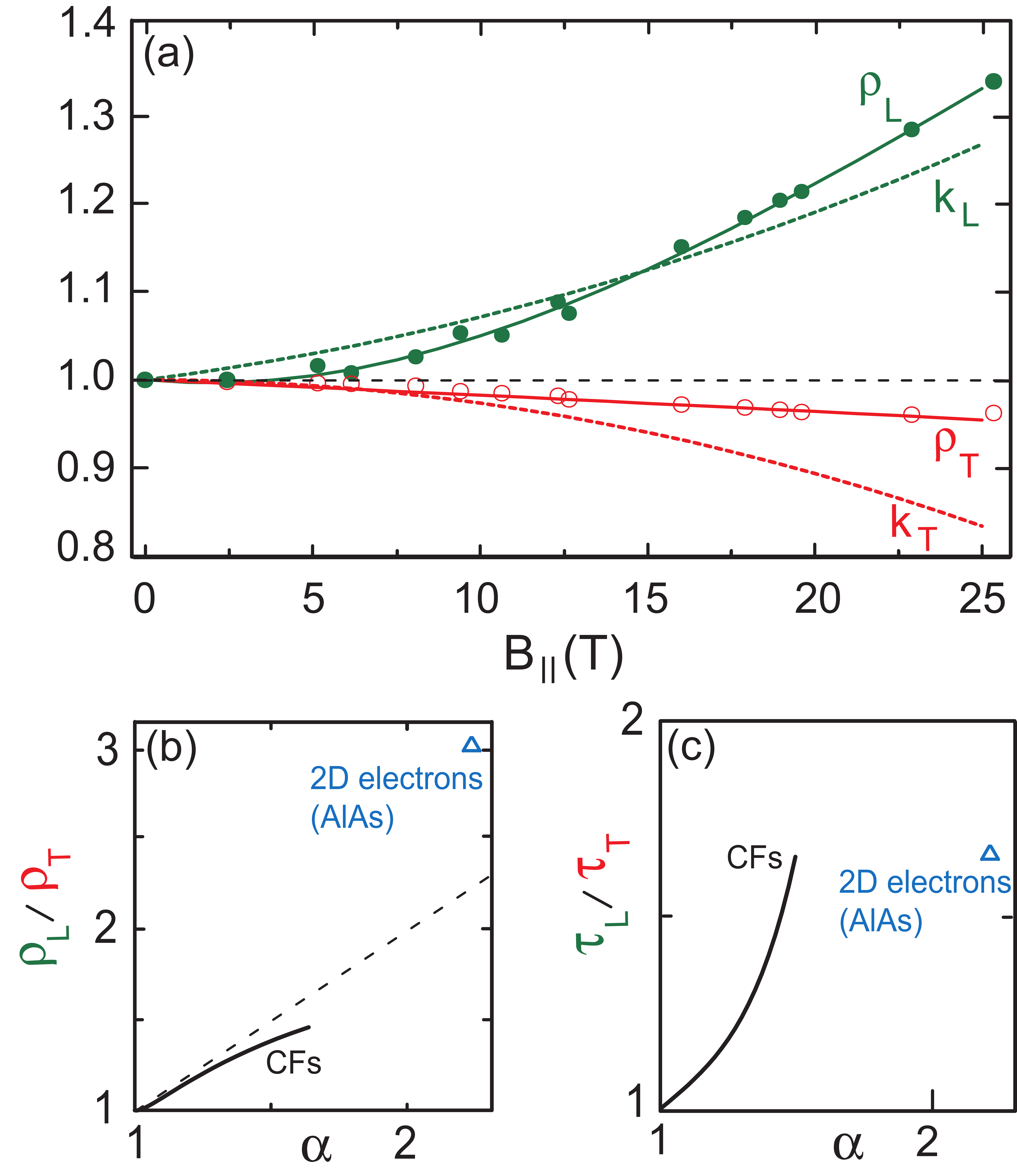}
\caption{\label{fig:Fig1} (color online) (a) CF resistivity at $\nu=1/2$ ($\rho_L$ and $\rho_T$) and Fermi wave vector ($k_L$ and $k_T$) as a function of $B_{||}$, normalized to their respective $B_{||}=0$ values. (b) Resistivity anisotropy ($\rho_L/\rho_T$) vs $\alpha$ (= $k_L/k_T$). The dashed line  with unit slope shows that $\rho_L/\rho_T$ is sub-linear in $\alpha$. Note that the Drude model with an isotropic $\tau$ would predict $\rho_L/\rho_T\sim\alpha^2$. (c) Scattering time anisotropy ($\tau_L/\tau_T$) as function of $\alpha$. For comparison with another anisotropic (elliptical) system, in (b) and (c) we show the corresponding points (blue triangles) for 2D electrons in AlAs.}
\end{figure}
  
In the absence of any theoretical model for CFs' transport, we analyze the data of Fig. 3 using the Drude model expression, $\rho={m}/pe^{2}\tau$, where $m$ is the CFs' effective mass and $\tau$ is their transport scattering time. We emphasize that the applicability of the Drude model to CFs is not known. In this model, mass anisotropy directly translates into anisotropy of $\rho$ \textit{if} $\tau$ is isotropic: $\rho_{L}/\rho_{T}={m_{L}/m_{T}}$. Moreover, if the Fermi contour is elliptical, the ratio of the effective masses along the principal directions should be proportional to the ratio of the respective wavevectors squared, i.e.,  ${m_{L}/m_{T}}={k_L^2/k_T^2}= \alpha^2$ (we denote the Fermi contour anisotropy, $k_L/k_T$, by $\alpha$). In the case of CFs subjected to $B_{||}$, the geometric mean of $k_L$ and $k_T$ normalized to $k_F^o$ indeed stays very close to unity,  suggesting that their Fermi contour is elliptical up to large $B_{||}$ \cite{Kamburov.PRL.2013}. This implies ${\rho_L/\rho_T}={k_L^2/k_T^2}=\alpha^2$ for CFs, a behavior which is clearly not observed in Figs. 1(a) and 3(b), suggesting that CFs' $\tau$ is anisotropic when their Fermi contour becomes anisotropic. To illustrate the anisotropy of $\tau$, in Fig. 3(c) we plot $\tau_L/\tau_T$ as a function of $\alpha$. Clearly $\tau_L>\tau_T$ when $k_L>k_T$, meaning that CFs scatter less along the longitudinal direction where their momentum $\hbar{k_{F}}$ is greater. To comment on this, we consider large-angle scattering (e.g. back-scattering) of CFs which contributes the most to resistance. In the event of back-scattering, the initial and final states on the Fermi contour are separated by $\pi$. For an elliptical Fermi contour, the required change in the wave vector ($\Delta k_{F}$) to back-scatter would be larger along the longitudinal direction than the transverse direction. According to the Born approximation, the scattering rate is proportional to the squared amplitude of the Fourier transform of the scattering potential \cite{Davies book}. For large $\Delta k_{F}$, the Fourier transform of the scattering potential decays rapidly \cite{Davies book}. As a result, the scattering probability along the longitudinal direction is expected to be smaller compared to the transverse direction; this is consistent with our observation.

Do other anisotropic systems also show similar scattering time anisotropy in light of the Drude model? To address this question, we first look at the zero-field counterparts of our hole-flux CFs, i.e., 2D holes. While holes' Fermi contours also become anisotropic when subjected to $B_{||}$, there are important differences. For example, unlike CFs, $\rho$ for 2D holes increases in $both$ longitudinal and transverse directions \cite{tutuc.PRL.2001}. The gradual spin polarization of holes by $B_{||}$ causes a reduction of screening which enhances scattering and increases $\rho$. Moreover, the 2D holes' Fermi contour rapidly distorts into non-elliptical shapes with increasing $B_{||}$ \cite{Kamburov1.PRB.2012}, making an estimation of transport mass and the applicability of the Drude model problematic. Qualitatively similar phenomena are also observed for 2D electron systems (2DESs) in GaAs \cite{Mueed.PRL.2015}. Because of these complications, we consider the 2DES confined to an AlAs QW whose Fermi contour is already elliptical without any $B_{||}$ \cite{Gokmen.NatPhy.2010}. For such a 2DES, $m_L/m_T\simeq5.1$ or, equivalently, $\alpha\simeq2.25$, and $\rho_L/\rho_T\simeq3$ \cite{Gokmen.NatPhy.2010}. According to the Drude model, this implies that $\tau_L>\tau_T$, similar to what we observe for anisotropic CFs. For a quantitative comparison, we show $\rho_L/\rho_T$ and $\tau_L/\tau_T$ for AlAs in Figs. 3(b) and (c). Interestingly, for comparable scattering time anisotropy, CFs require a much smaller $\alpha$ than 2D electrons in AlAs.

\begin{figure}
\includegraphics[width=.49\textwidth]{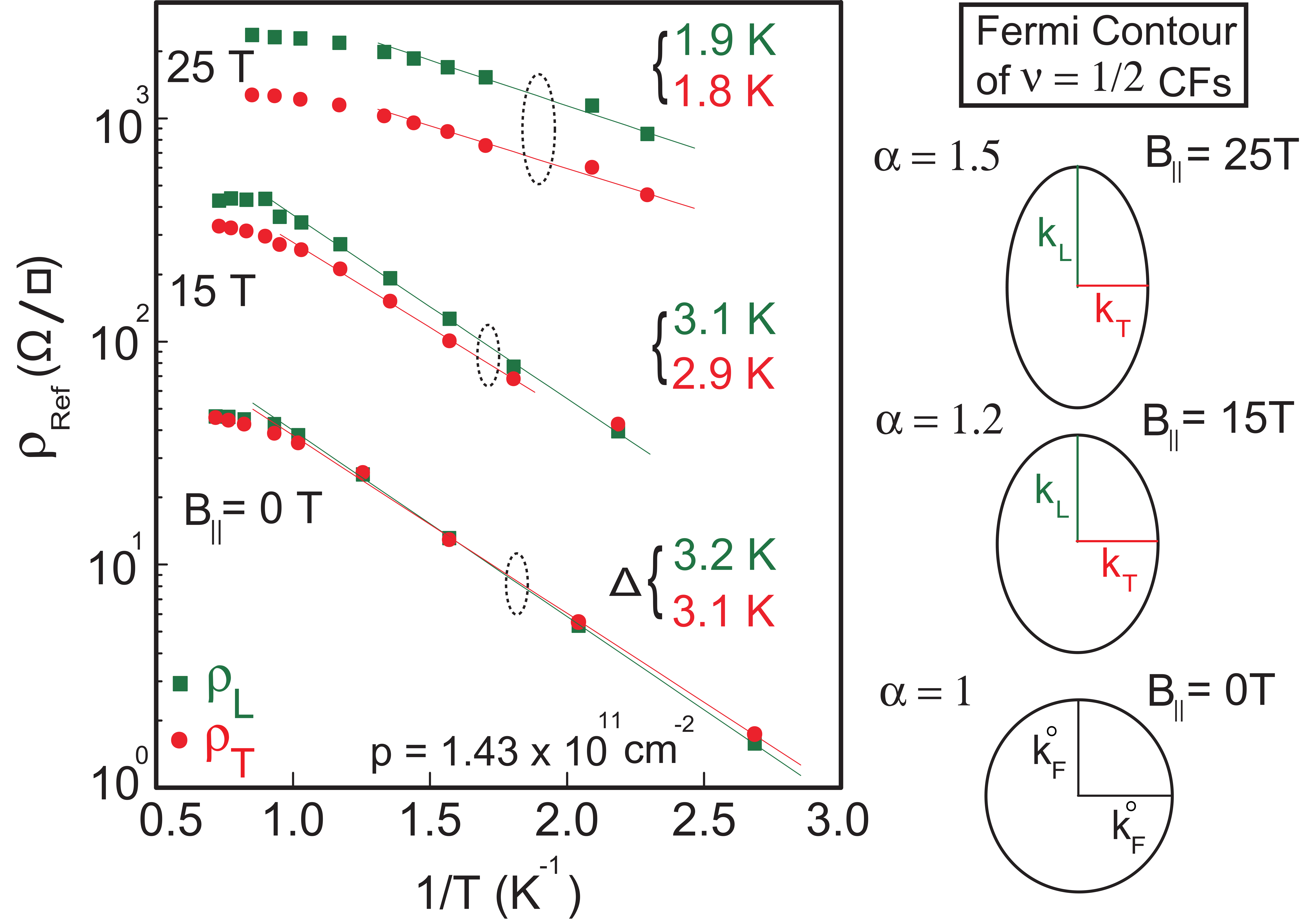}
\caption{\label{fig:Fig1} (color online) Arrhenius plots of $\nu=2/3$ FQHE resistivity minima along the transverse and longitudinal directions at three $B_{||}$ values. Each set of Arrhenius plots is shifted vertically for clarity. On the right side, the Fermi contours of $\nu=1/2$ CFs are shown at the corresponding $B_{||}$ values.}
\end{figure}

We next address the question whether the Fermi sea anisotropy of CFs affects the energy gap of the neighboring FQHE states. 
Figure 4 exhibits the Arrhenius plots of the $\nu=2/3$ FQHE resistivity minimum for the longitudinal and transverse directions, taken at three different $B_{||}$ values. On its right side we also show the experimentally measured Fermi contours of $\nu=1/2$ CFs \cite{Kamburov.PRL.2013}.
From the Arrhenius plots, it is clear that, similar to $\nu=1/2$ CFs, the $\nu=2/3$ resistivity becomes increasingly anisotropic as $B_{||}$ gets larger, with $\rho_L>\rho_T$. We also find that the $\nu=2/3$ FQHE energy gap ($\Delta$), deduced from the expression $\rho(T)\sim e^{-\Delta/2k_{B}T}$, remains fairly unchanged up to $B_{||}=15$ T but decreases by about 40$\%$ for $B_{||}=25$ T when $\alpha$ $(\simeq1.5)$ becomes significantly large (for a plot of $\Delta$ vs $B_{||}$, see Fig. 1(b)). Although the gap decreases, its values are essentially the same for the longitudinal and transverse directions, as expected \cite{footnote4}. 

It is tempting to hypothesize that the reduction of $\Delta$ might be related to the significant $B_{||}$-induced anisotropy in our 2DHS. Theoretical studies qualitatively corroborate our hypothesis. Laughlin's explanation of the odd-denominator FQHE states as  incompressible liquids of interacting particles \cite{Laughlin.PRL.1983} and most subsequent studies assume isotropy in the Coulomb interaction. However, the application of $B_{||}$ can induce interaction anisotropy which is closely linked to the Fermi sea anisotropy of the $\nu=1/2$ CFs \cite{Yang.PRB.2013}. In such an anisotropic system, Coulomb interaction is $\propto1/\sqrt{x^{2}{\alpha}+y^{2}/{\alpha}}$), where $x$ and $y$ are position coordinates along the principal axes of the Fermi contour \cite{Balagurov.PRB.2000, Wang.PRB.2012,Gokmen.NatPhy.2010,Yang.PRB.2013,Balram.preprint}. With increasing $\alpha$, the Coulomb interaction becomes increasingly one-dimensional. As a consequence, the isotropic 2D character of a FQHE state diminishes and its energy gap is therefore expected to decrease. Indeed, calculations indicate that while the FQHE states of the lowest Landau level are robust against moderate interaction anisotropy, their energy gaps decrease when anisotropy becomes substantial \cite{Bo.PRB.2012,Wang.PRB.2012,Balram.preprint,Papic.PRB.2013}.  Although we find qualitative agreement between our data and the calculations, the predicted reduction of $\Delta$ is smaller. For example, according to Ref. \cite{Balram.preprint}, $\alpha \sim1.5$ reduces $\Delta$ by a small amount ($<5\%$).


While we do not understand the reason for this discrepancy, we consider several other possibilities which could contribute to the reduction of $\Delta$: 

(i) \textit{Spin transition} {-}{-} It is well understood that $B_{||}$-induced FQHE spin-polarization transition in low-density 2DESs \cite{Kukushkin.PRL.1999, Liu.PRB.2014, eisenstein.PRB.1990,engel.PRB.1992} can reduce $\Delta$. However, we can rule it out since at the high density of our 2DHS, spin transitions are neither observed nor expected \cite{footnote2_1}. 

(ii) \textit{Single-layer to bilayer transition} {-}{-} Such a $B_{||}$-induced transition in the charge distribution, as observed in 2DESs confined to very wide QWs \cite{Mueed.PRL.2015, lay.PRB.1997, Hasdemir.PRB.2015}, can also reduce $\Delta$. As discussed earlier, in a quasi-2D carrier system, $B_{||}$ couples to the out-of-plane motion of the carriers. When the magnetic length corresponding to $B_{||}$, ($l_{B_{||}}$) becomes smaller than the QW width, the charge distribution transforms from a single-layer to bilayer \cite{Mueed.PRL.2015, lay.PRB.1997, Hasdemir.PRB.2015}. This transformation is reflected in the Fermi contour, too. The circular Fermi contour, at $B_{||}=0$, becomes distorted in the presence of $B_{||}$. The shape of the Fermi contour gradually evolves into a \textit{peanut} with increasing $B_{||}$, and eventually splits into two \textit{tear-drops} when the system becomes bilayer \cite{Mueed.PRL.2015}. We expect a qualitatively similar evolution for our 2DHS. At $B_{||}\simeq 25$ T, the magnetic length $l_{B_{||}}\simeq5$ nm is much smaller than the QW width (17.5 nm), and simulations \cite{Kamburov1.PRB.2012, R.L.Winkler.unpublished} indeed indicate that the Fermi contour attains a peanut shape and is on its way to split at very large $B_{||}$. However, this observation is for 2D holes at $B_{\perp}\sim0$ and does not necessarily reflect their charge distribution under large $B_\perp$ at filling factors such as $\nu=2/3$ or 1/2. Commensurability measurements of CFs (Fig. 2(d)) in fact confirm that, near $\nu=1/2$, our 2DHS stays single-layer like up to $B_{||}\simeq 25$ T \cite{Kamburov.PRL.2013}. One might conclude that, for the nearby $\nu=2/3$ FQHE, the charge distribution is also single-layer like under comparable $B_{||}$. However, there is the possibility that at $\nu=2/3$ the interacting 2DHS would prefer to form a bilayer charge distribution and host FQHE states at 1/3 fillings in each layer \cite{lay.PRB.1997}. Such a transition would result in a minimum in $\Delta$ vs $B_{||}$ \cite{lay.PRB.1997}, and could explain the large reduction in $\Delta$ we observe at large $B_{||}$. Unfortunately, we could not measure $\Delta$ at higher $B_{||}$ because of the maximum magnetic field available in our experiments. 

(iii) \textit{Disorder} {-}{-} Another possible scenario is that disorder plays a larger role at high $B_{||}$ and contributes to a reduction of $\Delta$. The observation of a significant $increase$ in $\Delta$ at high $B_{||}$ in hetero-structure 2DESs \cite{Haug.PRB.1987}, however, argues against this possibility. It is worth noting that, for a 2DES confined to a GaAs/AlGaAs hetero-structure, we do not expect any single-layer to bilayer transition at large $B_{||}$ because of its very narrow wave function thickness. For similar reason, we also expect negligible CF Fermi contour anisotropy \cite{Kamburov.PRB.2014}. The fact that $\Delta$ does not decrease in hetero-structure samples is consistent with our above discussion. The increase in $\Delta$, however, can be partly attributed to the narrowing of the wave function with increasing $B_{||}$ which enhances the electron-electron interaction \cite{Haug.PRB.1987, Shayegan.PRL.1990}. 

We also measured the energy gap of the $\nu=2/3$ FQHE for a 2DES confined to a 40-nm-wide GaAs QW with density $\simeq 1.75\times10^{11}$ cm$^{-2}$. The experiments revealed that $\Delta$ decreases from $\simeq2.3$ K at $B_{||}=0$ to $\simeq1.2$ K at $B_{||}=19.5$ T when $\alpha\simeq$ 1.5 for the $\nu=1/2$ CFs \cite{Kamburov.PRB.2014}. The relative change in $\Delta$ is close to that of the 2DHS sample. In this case, too, there is a possibility that the reduction in $\Delta$ is partly caused by the tendency of the 2DES to become bilayer at $\nu=2/3$ at very large $B_{||}$ even though the CF commensurability data indicate that near $\nu=1/2$ the system has a single-layer character \cite{Kamburov.PRB.2014}.

Data presented here demonstrate how the anisotropy of CFs' Fermi contour, tuned by a parallel magnetic field, affects their fundamental properties. Treating CFs using the Drude model reveals how the CF Fermi contour anisotropy affects their scattering time. The results also suggest that the energy gap for the $\nu=2/3$  FQHE decreases in the presence of large anisotropy, although we cannot rule out that the decrease is partly caused by a tendency of the 2DHS charge distribution towards a bilayer system at very large parallel fields. Our results should stimulate future theoretical studies to explore the transport properties of anisotropic CFs and FQHE.

We acknowledge support through the DOE BES (Grant DE-FG02-00-ER45841) and the NSF (Grant DMR-1305691) for measurements on 2D holes and 2D electrons, respectively. We also acknowledge the NSF (Grants ECCS-1508925 and MRSEC DMR-1420541), the Gordon and Betty Moore Foundation (Grant GBMF4420), and the Keck Foundation for sample fabrication and characterization, and  Our work was performed at the National High Magnetic Field Laboratory (NHMFL), which is supported by the NSF Cooperative Agreement DMR-1157490, by the State of Florida, and by the DOE. We thank J. K. Jain for illuminating discussions, and S. Hannahs, T. Murphy, A. Suslov, J. Park and G. Jones at NHMFL for technical support.


\begin{thebibliography}{99}
\bibitem{footnote3} In Fig. 2(d), the surface grating period for the black and green traces is 175 nm but for the red trace is 200 nm. The $B_{\perp}^{*}$-axis for the red trace has been rescaled by a factor of (200/175) to account for its different modulation period.
\bibitem{footnote1} The measured resistivity at $B_{||}=0$ along the $[\overline{1}10]$ arm is $\simeq6\%$ higher than that along the [110] arm. In order to simplify the analysis, we normalize the resistivity data for both directions to the corresponding values at $B_{||}=0$.
\end{thebibliography}
\end{document}